# Optimal Placement of Cores, Caches and Memory Controllers in Network On-Chip

Diman Zad Tootaghaj
dxz149@psu.edu

Farshid Farhat
fuf111@psu.edu

*Abstract*—**Parallel programming is emerging fast and intensive applications need more resources, so there is a huge demand for on-chip multiprocessors. Accessing L1 caches beside the cores are the fastest after registers but the size of private caches cannot increase because of design, cost and technology limits. Then split I-cache and D-cache are used with shared LLC (last level cache). For a unified shared LLC, bus interface is not scalable, and it seems that distributed shared LLC (DSLLC) is a better choice. Most of papers assume a distributed shared LLC beside each core in on-chip network. Many works assume that DSLLCs are placed in all cores; however, we will show that this design ignores the effect of traffic congestion in on-chip network. In fact, our work focuses on optimal placement of cores, DSLLCs and even memory controllers to minimize the expected latency based on traffic load in a mesh on-chip network with fixed number of cores and total cache capacity. We try to do some analytical modeling deriving intended cost function and then optimize the mean delay of the on-chip network communication. This work is supposed to be verified using some traffic patterns that are run on CSIM simulator.**

## I. INTRODUCTION

The purpose of optimal cache placement design is to reduce the average latency, assuming the same amount of total aggregated capacity. Increasing the total aggregated capacity would reduce total number of memory requests. At the same time the optimal design should take into account the effect of network hop delay. The ever increasing on-chip communication leads to the designs where each core has its own private cache to reduce the congestion in the network. However, the working sets of applications is also increasing, having private caches at each core for shared data is not efficient because each cached block should be replicated in all private caches.

## II. THE PLACEMENT OF CORES, CACHES AND MEMORY CONTROLLERS

### A. Assumptions

Having a $n \times n$ mesh on-chip interconnection network, we are interested in finding where to place cores, caches and memory controllers to optimize expected access latency of all cores. We assume that each core has a private L1 cache, L1 and L2 caches are exclusive, and L2 caches are shared among all blocks. The misses in L2 cache are directly sent to the memory controllers. The number of cores ($N_r$) is assumed to be fixed and the total last level cache capacity is also constant or we can say the number of fixed-size caches is $N_h$. Also number of memory controllers ($N_m$) is fixed.

Mesh on-chip network is a 2D plane with a fixed area (big tile) where a core fills a fraction of it as a small tile that cannot exceed to other tiles of cores or caches. We assume the probability of having a hit in L2j for a request from $core_i$ is $p_{ij}$, which is called $core_i$'s hit ratio at L2j. In this work we'll consider mesh interconnection networks up to 16x16. "NoC configs" in the following equations are all $\binom{N_r + N_h + N_m}{N_r, N_h, N_m}$ possible combinations of the placement of the cores, caches, and memory controllers in a mesh on-chip network, where the number of combinations is growing exponentially (why?).

## III. PROBLEM DEFINITION

The goal is to minimize the expected access latency of the cores to the all blocks in all levels of memory hierarchy as:

$$\min_{NoC\ configs} \sum_{i=1}^{N_r} E\{Latency_{core_i}\}$$

We have the hit and miss ratios of $core_i$ to each block of $cache_j$, and we assume that the probability of access to $cache_j$ is equal with the probability of access to $cache_k$, so for the expected latency of $core_i$ we have:

$$\begin{aligned}E\{Latency_{core_i}\} &= Latency_{L1} * HitRatio_{L1} \\ &+ (Latency_{L1} + Latency_{L2}) \\ &\quad * MissRatio_{L1} * HitRatio_{L2} \\ &+ (Latency_{L1} + Latency_{L2} \\ &\quad + Latency_{Memory}) * MissRatio_{L1} \\ &\quad * MissRatio_{L2} * HitRatio_{Memory}\end{aligned}$$

In the above formula $Latency_{L1}$ is constant for all cores and $HitRatio_{Mem} = 1$, because we assume no load/store instruction goes to the disk, and then we have:

$$\begin{aligned}&E\{Latency_{core_i}\} \\ &= Latency_{L1} * (HitRatio_{L1} + MissRatio_{L1} * HitRatio_{L2} \\ &\quad + MissRatio_{L1} * MissRatio_{L2}) \\ &\quad + Latency_{L2} \\ &\quad * (MissRatio_{L1} * HitRatio_{L2} \\ &\quad + MissRatio_{L1} * MissRatio_{L2}) \\ &\quad + Latency_{Memory} * MissRatio_{L1} \\ &\quad * MissRatio_{L2} \\ &= Latency_{L1} + Latency_{L2} * MissRatio_{L1} + Latency_{Mem} \\ &\quad * MissRatio_{L1} * MissRatio_{L2}\end{aligned}$$

$Latency_{L1}$, $MissRatio_{L1}$ and $MissRatio_{L2}$ are not changed in different placements because the total cache size and memory mapping on the caches are the same for all configurations.

*1) Low Traffic*

In the case of low traffic $Latency_{L2}$ depends on the Manhattan distance of $core_i$ to $cache_j$, and $Latency_{Mem}$ depends on the Manhattan distance of $cache_j$ to $memory-controller_k$, and the expected latencies are:

$$Latency_{L2_{core_i}} \approx E_i\left\{\left\|X_{core_i} - X_{cache_j}\right\|_1\right\}$$
$$= \sum_{j=1}^{N_h} p_{ij} \left\|X_{core_i} - X_{cache_j}\right\|_1$$
$$Latency_{Mem_{core_i}} \approx E_i\{E_j\{\left\|X_{cache_j} - X_{mc_k}\right\|_1\}\}$$

Based on these expected latencies, the minimization problem can be simplified to:

$$\min_{NoC\ configs} \sum_{i=1}^{N_r} E\{Latency_{core_i}\}$$
$$= \min_{NoC\ Configs} \sum_{i=1}^{N_r} \{Latency_{L2_{core_i}} * MissRatio_{L1} + Latency_{Mem_{core_i}} * MissRatio_{L1} * MissRatio_{L2}\}$$

Because of the separability of optimal cache placement and optimal memory-controller placement problems, we are interested to find the optimal placement of cores and caches in the defined problem as follows:

$$\min_{NoC\ Configs} \sum_{i=1}^{N_r} Latency_{L2_{core_i}}$$
$$= \min_{NoC\ Configs} \sum_{i=1}^{N_r} \sum_{j=1}^{N_h} p_{ij} \left\|X_{core_i} - X_{cache_j}\right\|_1$$

It is a L1-norm optimization problem constrained to possible NoC configurations.

*2) High Traffic*

In the case of high traffic $Latency_{L2}$ depends on the routing path from $core_{(x_i,y_i)}$ to $cache_{(x_j,y_j)}$ so response time (RT) of each intermediate router matters, and $Latency_{Mem}$ depends on the routing path of $cache_{(x_j,y_j)}$ to $memory-controller_{(x_k,y_k)}$, so response time (RT) of each passed router in the path should be included, so by assuming static XY-routing the expected latencies are:

$$Latency_{L2_{core_{(x_i,y_i)}}} = E_i\left\{\left\|core_{(x_i,y_i)}, cache_{(x_j,y_j)}\right\|_1\right\}$$
$$= E_i\left\{\sum_{k=i\ @\ y_i}^{j} RT_{router_{(x_k,y_i)}} + \sum_{k=i\ @\ x_j}^{j} RT_{router_{(x_j,y_k)}}\right\}$$
$$= \sum_{j=1}^{N_h} p_{ij}\left\{\sum_{k=i\ @\ y_i}^{j} RT_{router_{(x_k,y_i)}} + \sum_{k=i\ @\ x_j}^{j} RT_{router_{(x_j,y_k)}}\right\}$$
$$Latency_{Mem_{core_{(x_i,y_i)}}} = E_i\{E_j\{\left\|cache_{(x_j,y_j)}, mc_{(x_k,y_k)}\right\|_1\}\}$$

Based on these expected latencies, the minimization problem can be simplified to:

$$\min_{NoC\ configs} \sum_{i=1}^{N_r} E\left\{Latency_{core_{(x_i,y_i)}}\right\}$$
$$= \min_{NoC\ Configs} \sum_{i=1}^{N_r} \{Latency_{L2_{core_{(x_i,y_i)}}} * MissRatio_{L1} + Latency_{Mem_{core_{(x_i,y_i)}}} * MissRatio_{L1} * MissRatio_{L2}\}$$

Now we are interested to find the optimal placement of cores and caches in the defined problem as follows:

$$\min_{NoC\ Configs} \sum_{i=1}^{N_r} Latency_{L2_{core_{(x_i,y_i)}}}$$
$$= \min_{NoC\ Configs} \sum_{i=1}^{N_r} \sum_{j=1}^{N_h} p_{ij} \left\|core_{(x_i,y_i)}, cache_{(x_j,y_j)}\right\|_1$$

**Lemma 1:** If the cores and caches were already placed, there would be at least one optimal placement of the memory controllers in a mesh on-chip network with the defined assumptions. And the minimization problem is:

$$\min_{NoC\ Configs} \sum_{i=1}^{N_r} Latency_{Mem_{core_{(x_i,y_i)}}}$$

**Lemma 2:** If the caches are already placed, there is at least one optimal placement of the cores in a mesh on-chip network with the defined assumptions. And the optimization problem is:

$$\min_{NoC\ Configs} \sum_{i=1}^{N_r} Latency_{L2_{core_{(x_i,y_i)}}}$$

Optimal placement of memory controllers can be solved by any parametric configuration of cores and caches, because it just depends on the parametric locations of the caches in the NoC and memory controllers can be placed off-chip without area issue.

**Corollary 3:** Placement of core, caches and memory controllers in a NoC can be broken into two optimization problems that the first one wants to optimize the cores and caches placement, and the other one wants to find the optimized locations of memory controllers based on optimal locations of the cores and caches.

## IV. MODELING OF RESPONSE TIME IN ROUTERS OF MESH NETWORK-ON-CHIP

In this section it is tried to model response time in routers of mesh network on chip. In previous section we found a closed formula of expected delay of an end to end message with respect to response time of intermediate routers, now in this section we focus on derivation of a formula for response time of mesh NoC routers. In the literature there are so many analytical modeling of NoC routers but for symmetric networks like hypercube [10, 13]. Mesh network nature is asymmetric with respect to all routers (only routers with the same distance from center are the same). There are a few papers pertaining to analytical performance model of mesh network but they tried to approximate or ignore channel contention by fully adaption routing [11], enough virtual channels [12], or assuming prioritized queues [14, 20, 21].

These assumptions are rigid and make the model far from the reality or even faulty.

Hop (router) delay or response time can be divided into two parts: deterministic part and stochastic part. Deterministic part of delay is the all-time fixed delay of sending a packet of size $L$ from A to B, when the parameters (A,B,L) are fixed for a given NoC. This part is equivalent with the very low traffic section where we were investigated some results. Now we are interested in formulating the stochastic part as well. Stochastic part of packet latency is related to the total waiting time of the packet in intermediate buffers.

Each router has $n$ physical channels and virtual cut-through switching under XY-routing algorithm is executed in mesh NoC. The size of the packet is an exponential random variable ($L$), and packet generation rate is $\lambda_g$ uniformly on all destined caches. Residual packet waiting time is shown by $R$, and $S_i; i = 1..n$ shows the service time of a packet as a random variable where $E\{S_i\} = 1/\mu$.

Algorithm 1 (Packet delay inspector): It is assumed that mesh NoC is given. The packet delay inspector algorithm works as follows:

1 Based on the application specification, get the packet generation rates ($\lambda_g$ s) of each core to the caches, or as a general purpose solution assume it is constant and uniform over all caches.

2 Given packet generation rates, topology, and NoC routing algorithm we can find the inter-arrival rates ($\lambda_i; i = 1..n$) to the input channels of intermediate routers. Because given inter-arrival rates to some input channels and the destination of the rates we can find the output channels rates (inter-departure rates) or equivalently input channel rates of the next hops. In steady state by ignoring contention we can find output channels rates (why? As a lemma it should be shown).

3 Given inter-arrival rates to input channels and the destination of the rates in a router, the contention matrix of the router can be found.

4 Given contention matrix, inter-arrival rates and service rates of the intermediate routers, any packet delay from some A to some B can be calculated. In fact this delay in steady state is the summation of expected (mean) delay of all intermediate routers in the path of the packet.

First and second parts are straightforward and we are going to explain the rest. Effective utilization matrix with respect to service time, inter-arrival rate matrix and contention matrix is:

$$\rho_e = \lambda C. E\{S\}; \; S = \begin{bmatrix} S_1 & 0 & 0 & 0 \\ 0 & S_2 & 0 & 0 \\ 0 & 0 & ... & 0 \\ 0 & 0 & 0 & S_n \end{bmatrix}_{n \times n}, \lambda = \begin{bmatrix} \lambda_1 & 0 & 0 & 0 \\ 0 & \lambda_2 & 0 & 0 \\ 0 & 0 & ... & 0 \\ 0 & 0 & 0 & \lambda_n \end{bmatrix}_{n \times n}, C = [c_{ij}]_{n \times n}$$

Now based on Kingsman's formula we can write the waiting time as follows:

$$W_q \approx \frac{E\{R\}}{1 - \rho_e} = \frac{C_a^2 + C_s^2}{2} \frac{E\{S\}}{1 - \rho_e}$$

where $C_a$ is the coefficient variation of arrivals and $C_s$ is the coefficient variation of services. Now only we have to specify the $c_{ij}$ elements of contention matrix. If $N_j$ is the length of j-th input queue, we have:

$$c_{ij} N_j = \rho_i \Pr(\lambda_{ik} @head) \sum_{m=1}^{\infty} m \left( \rho_j \Pr(\lambda_{jk} @position\ m) \right)^m$$

For example in case of M/G/1 $\Pr(\lambda_{jk} @position\ m) = 1 - e^{-\lambda_{jk} E\{S_j\}}$.

## V. SIMULATION RESULTS

In this section we describe the simulation environment and setups. In order to find service time, utilization, throughput queue length, and response time of different routers we use CSIM simulator. We use cut-through switching under x-y routing which is deadlock free for a mesh network. Message size is an exponential random variable with the mean of 10, i.e., on average each message contains ten packets going through the buffers of the routers. The total service rate of all routers is assumed a constant value for all configurations. Simulation runs for 100,000 times for a particular cache-core configuration to find the reliable statistics. Figure 1. Shows different cached configurations we used during the simulation. Figure 1.a shows a cached configuration where all caches are in the middle and 1.e shows a fully distributed configuration. We assume the total aggregated cache capacity is fixed. We observed that for common traffic load the centralized configuration fig1.a has smaller average message latency compared to other configurations. However by increasing the traffic load other configurations respectively Fig.1(b), Fig.1(c), and Fig.1(d) may become better and finally fully distributed cache, fig.1.e has better performance in high load. Figure 2-6 shows different statistics for the routers in the diagonal of the mesh network configurations in Figure1. Because of symmetry we didn't bring the results for all routers.

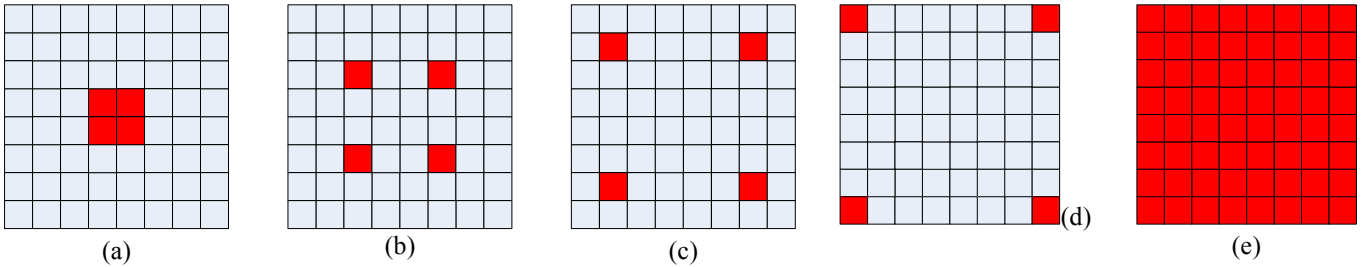

(a) (b) (c) (d) (e)

Fig. 1. Different cache configurations used in simulation.

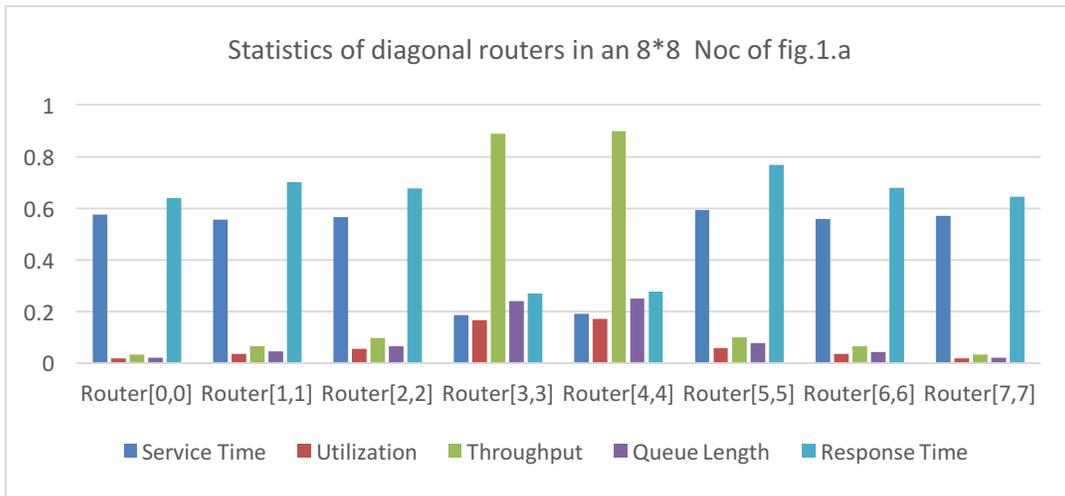

Fig. 2. Service time, Utilization, Throughput, Queue Length and response time for the diagonal routers in an 8*8 mesh Noc network of fig1.a.

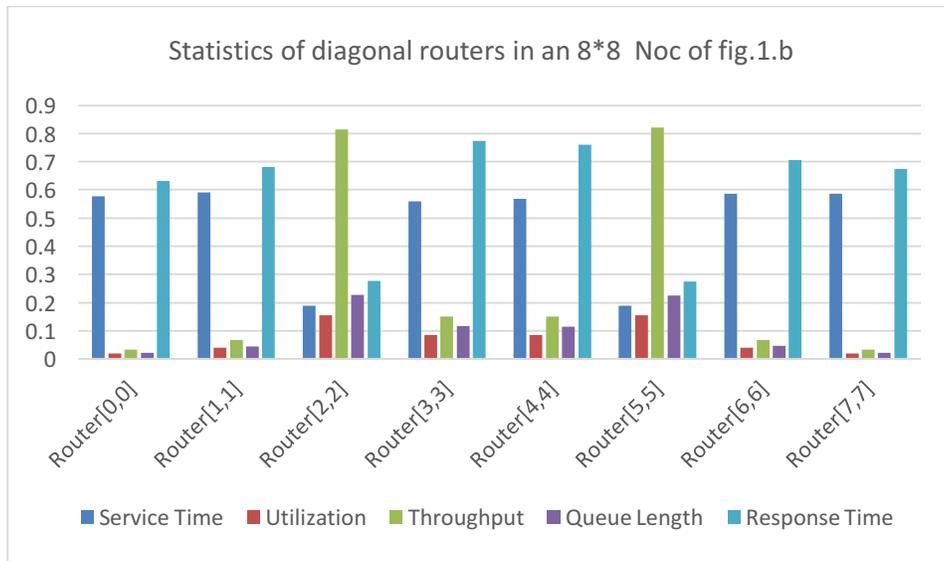

Fig. 3. Service time, Utilization, Throughput, Queue Length and response time for the diagonal routers in an 8*8 mesh Noc network of fig1.b.

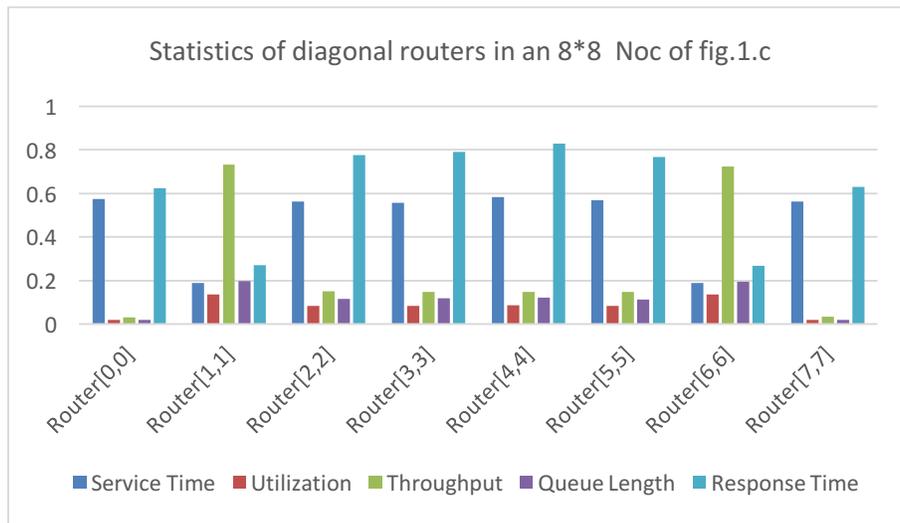

Fig. 4. Service time, Utilization, Throughput, Queue Length and response time for the diagonal routers in an 8*8 mesh Noc network of fig1.c.

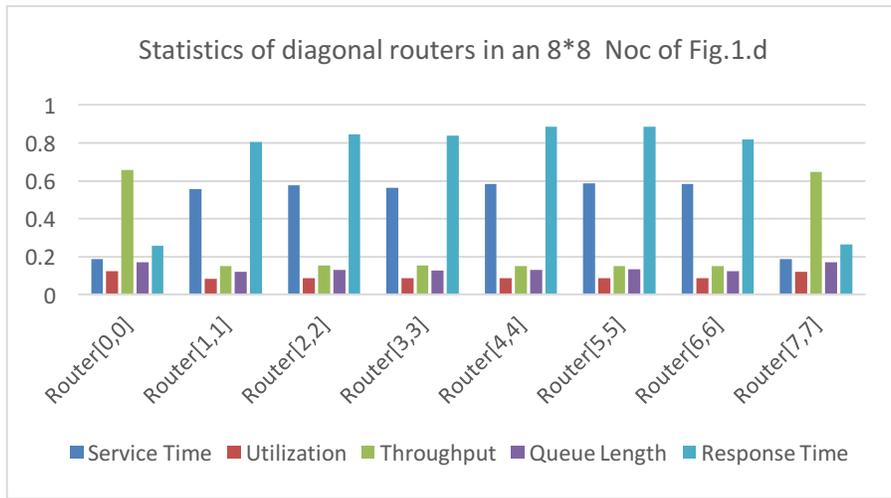

Fig. 5. Service time, Utilization, Throughput, Queue Length and response time for the diagonal routers in an 8*8 mesh Noc network of fig1.d.

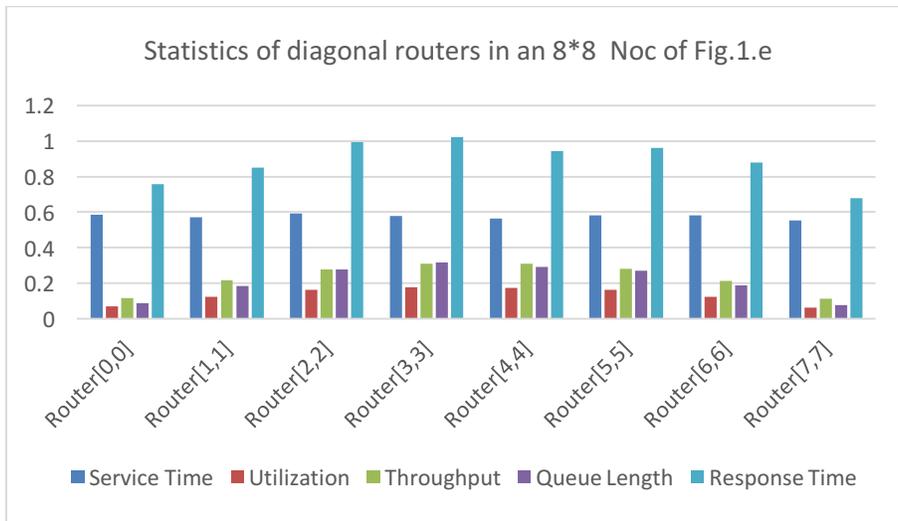

Fig. 6. Service time, Utilization, Throughput, Queue Length and response time for the diagonal routers in an 8*8 mesh Noc network of fig1.e.

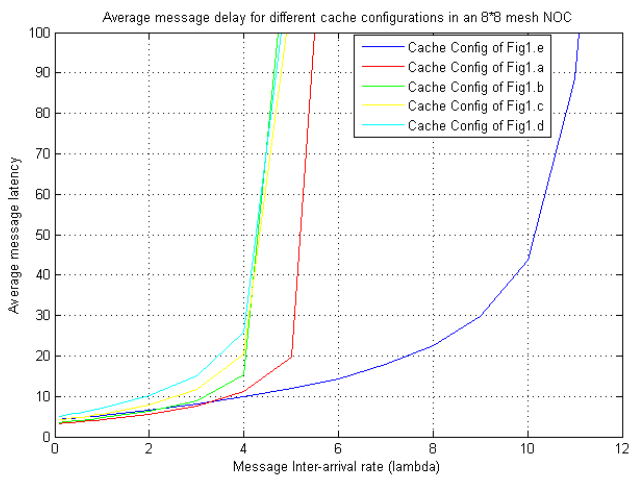

Fig. 7. Average message latency for different inter-arrival rates of different cache configurations.

Figure 7 shows average message latency for different cache configurations of Figure 1. It was observed that for common inter-arrival times in a mesh Noc (smaller than 3) the central cache configuration performs better than the other configurations. However for high traffic loads distributed approach has better performance and reaches the saturation point later.

## VI. RELATED WORKS

Researches on optimal placement of cores, caches and memory controllers in NoC can be divided into three categories. The first category looks in how to place memory controllers to reduce the traffic contention. The second category tries to exploit heterogeneity or smart data placement to get advantage of locality and the third group tries to find best cache placement to reduce access latencies. In this project we are interested in finding a hardware design using mathematical models to find the best placement which reduces access latency of each block and we want to evaluate the theoretical model using experiments.

## A. Memory Controller placement

Abts et al. [1] have examined the placement of memory controllers in a mesh and torus based NoC to reduce the contention in the network and maximize delivered bandwidth. They assume having a fixed number of memory controllers, which is smaller than the number of cores, and each processor chooses one of memory controllers randomly. But it does not seem perfect, because they investigated some limited heuristic cases without any analytical support. They also haven't considered cache placement problem, however we know that cache placement and memory controller placement are two sides of the same coin and each one affects the other. In addition the authors haven't examined all possible placements because having $m$ memory controllers and $n$ cores the complexity of checking all possibilities is $\binom{n}{m}$. In order to reduce the complexity of checking all possibilities for large networks ($n > 6$), genetic algorithm and random walk is used. So the proposed placement is near-optimal. Also the proposed solution uses maximum channel load as a metric to compare network congestion for different configurations, and maximum channel load is computed by associating a counter to each link and tracking each packets traversal in the network, once a packet passes a link they increment the counter of that link. However maximum channel bandwidth should be compared with the maximum bandwidth of that link, also simply associating a counter to each link and comparing the maximum value of these counters is not always a good metric. As we will explain, the response time of each link doesn't increment linearly with $\lambda$, this assumption is true, only when $\lambda$ is very small with respect to service rate. Mean response time of a single M/M/1 queue is:

$$mean\ response\ time = \frac{1}{\mu - \lambda} = \frac{1}{\mu\left(1 - \frac{\lambda}{\mu}\right)}$$
$$= \frac{1}{\mu}(1 + \rho + \rho^2 + \cdots)$$
$$\approx \frac{1}{\mu}(1 + \rho)\ \ when\ \lambda\ is\ small$$

Farhat et al. [15, 16, 17, 22] studied a queueing theory analysis of Mapreduce jobs for large scale data computing jobs. In [18, 19] the authors study the affect of memory, cores and cache latency on the performance of high performance computing (HPC) applications.

In this project we want to consider the effect of both high load and low load traffic on the placement algorithm. After finding near-optimal memory controller placement, the paper investigates different routing algorithms for the placements which was found using exhaustive search or random simulation. They examined X-Y, Y-X, Y-X and X-Y routing algorithms for a set of placements and propose a new routing algorithm which performs better than the former routing algorithms. But memory controller placement and routing are not two independent problems and the optimal placement depends on the routing. In this project we want to use a mathematical model to solve the placement of cores, caches and memory controllers. To verify the mathematical model we'll use simulation model to investigate the traffic pattern and we'll use X-Y routing and find the contention of each link for a set of configurations. So for simulation part we won't examine all possible placements but it can be used to verify the analytical results.

## B. Data Placement and Heterogeneity

In [2], the authors have demonstrated using asymmetric-cache CMP design and effective asymmetric-aware and contention-aware scheduling support, delivers higher performance for each watt than the symmetric approach. It is shown that different applications need different cache size and for some applications increasing cache size doesn't help to improve the performance since they have very small working sets or because the benchmark is streaming so it does not benefit from larger cache size. This approach uses a system which has 4 cores and 2 cores share a big LLC and the other two cores share smaller LLC. This approach works well if the system is going to run a mix of applications which need big cache and small cache. However we have a lot of possible combinations of different applications so using this method is very limited. Also since this approach is somehow orthogonal to our project because in this project we are interested in a system of many cores NoC where all cores share the last level cache and we want to find where to place the cores. Also it is possible and reasonable that heterogeneity exists not only in cache level but also in core level, so in this case it is possible that some benchmarks need larger cache but smaller cores, so designing these kind of heterogeneous systems would be more complicated.

Hardavellas et al. [3] proposed a distributed cache design where modifiable blocks are mapped to only one location to reduce the overhead of coherency mechanism. Read-only blocks can be replicated in distant cores. The paper suggest that as the working sets of applications are growing, private cache architectures are inefficient because cache blocks are being replicated in different places. In addition a shared cache design that maps cache blocks to a fixed location is not efficient because some cache blocks would be accessed to distant caches frequently. The paper suggests that instead of placing the data randomly over all shared caches, replicating the shared read-only data to nearby cores. This paper is trying to solve the same problem that we want to solve in this project, but in a different dimension. In fact we want to find a design that reduces cache block access latency if we have static mapping of blocks. But in [3] the authors use the operating system to detect private and shared data and replicate the read-only shared data in the nearby locations to reduce access latency of distant blocks. Since only read-only blocks can be replicated they don't need to worry about coherency issues. We also don't want to consider coherency issues and we assume each block is mapped to only one location in the distributed last level caches. But for a fixed size aggregated capacity we want to see which placement would lead to minimum access latency considering network traffic.

In [8], the authors propose a heterogeneous network using small routers which are more power efficient and big routers which are high performance. The paper shows that the center of a mesh usually gets more congested than the corners or peripheral routers for a random and uniform traffic. The authors have also examined different memory controller placement configurations which was near optimal in [1] to verify their proposed design. But we should note that the traffic pattern depends on the cache, core and memory controller placement. So even though it is obvious that in a random and uniform traffic pattern the central routers buffer and link utilization is higher but this pattern depends both on the routing algorithm and how we place different components in the system.

*C. Cache Placement*

The only paper that directly wants to solve this problem is [4] by Xu et al. which explores the optimal core and cache placement problem for CMPs. They assume the area and number of cores and 30MB LLCs are equal. Their analytical model is based on average hop count accessing all LLCs for a core. They investigate ring-based and mesh-based interconnect networks where number of cores/caches is a multiple of four (4xN). From their formulation it can be inferred that core access to caches is randomly uniform. Then they try to solve the problem by an exhaustive search algorithm, so they manage to find some solutions for some limited number of configurations. They compare their optimal configurations with ring-based configuration using simulation.

That Xu et al. [4] do not consider channel contention and network congestion in their model at all; it is the main problem of the paper making it non-comprehensive. Their objective metric for minimization is very simple as a summation of the distance (hop-count) of each core to each cache, so the solution is sub-optimal and cannot be generalized to all practical situations. To get the optimized constellation they execute their exhaustive search algorithm and then based on the metric it would find the best constellation. Because the complexity of the problem is exponential, their algorithm can only find some small NoC configurations (4x4, 4x6, and 4x7). Equal number of cores and caches is not a rational assumption when a tile of a LLC cell has 30MB size versus a tile of a core cell with the same area. Based on their metric there are some other constellations equivalent with their solutions, which is not mentioned in paper. Astonishingly they just show their solutions' speedup versus ring-based NoC performance, and they don't compare with other mesh configurations but it is obvious that mesh-based NoC is better than ring-based.

Also Xu et al. [7] consider the optimal placement Through Silicon Vias (TSVs) in 3D NoC. Their analytical metric supposed to be optimized is the same as average hop count, and only the cores in lower layer access to the caches in upper layer via pillars. They just assess 8 or 16 pillar for 8x8 NoC because of high cost of manufacturing TSVs and their brute force algorithm cannot check all possible combinations because of high complexity of the problem. They show reducing number of pillars to quarter and 1/8 degrades the performance respectively 7% and 14% but we have a huge saving in the cost of manufacturing. Without considering NoC traffic pattern the solution is not valid. Also the same critiques of the previous paper remain for this paper.

*1) Centrally Placed Caches*

Liu et al. [5] show that the size of shared data among multi-threaded programs is low and it is better to keep shared part of L2 cache in the center cell of CMPs. In this paper 22 multi-threaded benchmarks including NAS, SpecOMP, Apache and SpecJbb are characterized by their shared L2 cache access pattern. The results show that the amount of active shared data to write is low but the number of accesses is high, so it's better to put the shared L2 cache near to all cores. As a result, the performance and power consumption would be improved. However the center cell cache seems rational, the work doesn't try to show the optimality of the design by any analytical modeling. The assumed interconnection network is ring that is not scalable and the chip layout has only 4 cores and 16MB L2 which 64KB kept in the middle for active writes. Nowadays NoC has many cells to place cores, caches and routers, and this work is limited to a special small design configuration. Consequently this work only shows that this approach makes sense but the general solution of cache placement problem still remains ambiguous.

A central cache design has been verified with better performance [6] versus distributed shared cache design that the cores are around the shared caches in the center of NoC. In order to access shared L2 cache in NoC, multiple ports with high bandwidth is necessary. Multi-threaded applications have multiple working sets to be put on limited on-chip caches. Shared blocks migration makes one processor's latency low but the others higher. They verify block migration inefficiency using full-system simulation, because near half of shared LLCs' hits happen in the banks which are in the center of mass of CMP. To manage wire delay for CMPs they apply a hybrid design including low-latency transmission lines and stride-based prefetching among L1 and L2. This work manages wire delay to access central caches very well, however it is restricted to CMPs without routers and it could not be extended to bigger NoC design where a scalable interconnection network is mandatory. When we enter the mesh-based NoC with routers, we encounter other issues like channel contention and network congestion which are not the investigated cases in this work.

## VII. CONCLUSION AND FUTURE WORKS

In this report we investigated the problem of cores, caches and memory controller placement in mesh NoC domain. The results show the effect of constellation and NoC traffic on the optimal solution. Based on the estimated traffic pattern symmetric optimal solution can be found with lower complexity. With regarding the derived results following works can be investigated in future works:

1. Investigating optimal placement for a more general accurate model of NoC.

2. Try to match the model with high load condition and consider fully, semi-fully, and non-uniform distributed cache placement.

3. There is an optimization problem of variable-rate routers placement given by fixed budget to achieve a better performance.

4. Break the main problem into easier problems like optimal memory-controller placement in case of fully distributed or other configurations.

5. Memory controllers shape the traffic a little as well, but these changes do not change the best placement of the joint problem mainly, and MCs can be placed inside the free region of caches symmetrically.